\documentstyle[11pt,aasms]{article}

\input epsf.sty
\tighten

\begin{document}
\title{DEGREE SCALE ANISOTROPY: SP94 RESULTS}

\author{J. O. Gundersen\altaffilmark{1,2}, M. Lim\altaffilmark{1,2}, J.
Staren\altaffilmark{1,2},
C. A. Wuensche\altaffilmark{3}, N. Figueiredo\altaffilmark{1,3,5}, T. C.
Gaier\altaffilmark{1,2},
T. Koch\altaffilmark{4}, P. R. Meinhold\altaffilmark{1,2},
M. D. Seiffert\altaffilmark{1,2}, G. Cook\altaffilmark{1}, A.
Segale\altaffilmark{1},
P. M. Lubin\altaffilmark{1,2}}
\slugcomment{Submitted to Astrophysical Journal Letters}
\altaffiltext{1}{Department of Physics, University of California, Santa
Barbara, CA 93106}

\altaffiltext{2}{NSF Center for Particle Astrophysics, University of
California,
Berkeley, CA 94720}

\altaffiltext{3}{Instituto Nacional de Pesquisas Espaciais-INPE/MCT, Divisao de
Astrofisica, Sao Jose dos Campos, SP, Brasil 12227-010}

\altaffiltext{4}{Jet Propulsion Laboratory, California Institute of Technology,
4800 Oak
Grove Drive, Pasadena, CA 91109}

\altaffiltext{5}{Escola Federal de Engenharia de Itajuba, Departamento de
Fisica e Quimica, Itajuba, MG, Brazil 37500-000}

\begin{abstract}
We present results from two observations of the cosmic microwave background
(CMB) performed from the South Pole during the 1993-1994 austral summer.
Each observation employed a 3$^{\circ }$ peak to peak sinusoidal, single
difference chop and consisted of a $20^{\circ }\times 1^{\circ }$ strip on
the sky near our SP91 observations. The first observation used a receiver
which operates in 3 bands between 38 and 45 GHz (Q-band) with a FWHM beam
which varies from $1^{\circ }$ to $1_{\cdot }^{\circ }15$. The second
observation overlapped the first observation and used a receiver which
operates in 4 bands between 26 and 36 GHz (Ka-band) with a FWHM beam which
varies from $1_{\cdot }^{\circ }25$ to $1_{\cdot }^{\circ }7$. The Ka-band
system has a similar beamsize and frequency coverage as the system used in
our SP91 results. Significant correlated structure is observed in all bands
for each observation. The spectrum of the structure is consistent with a CMB
spectrum and is formally inconsistent with diffuse synchrotron and free-free
emission at the 5 $\sigma $ level. The amplitude of the structure is
inconsistent with 20 K interstellar dust; however, the data do not
discriminate against flat or inverted spectrum point sources. The root mean
square amplitude ($\pm 1\sigma $) of the combined (Ka+Q) data is $\Delta
T_{rms}=42.0_{-6.8}^{+15.8}$ $\mu $K for an average window function which
has a peak value of $0.97$ at $\ell =68$ and drops to $e^{-0.5}$ of the peak
value at $\ell =36$ and $\ell =106$. A band power estimate of the CMB power
spectrum, $C_\ell $, gives $\left\langle \frac{C_\ell \ell (\ell +1)}{2\pi }%
\right\rangle _B=1.77_{-0.54}^{+1.58}\times 10^{-10}$. The band power
estimates for the individual Ka and Q-band results are larger than but
consistent with the band power estimate of the combined Ka-band SP91 results.
\end{abstract}


\clearpage

\section{Introduction}

Anisotropy measurements of the cosmic microwave background (CMB) are a very
effective tool for testing and constraining models of cosmic structure
formation. With the discovery of large angular scale anisotropy by COBE
(Smoot et al., 1992), there has been an increased interest in characterizing
anisotropy on degree angular scales. Although all CMB measurements have to
overcome a long list of systematic effects and foreground contaminants
(Wilkinson, 1994), they have the potential to constrain many of the global
parameters of the Universe and thus discriminate among the plethora of
cosmic structure formation models. Over the past six years, we have
travelled to the South Pole three times to perform these degree scale
anisotropy measurements. The results from our 1988-89 measurements are
detailed in Meinhold and Lubin, 1991 (SP89a) and Meinhold et al., 1993
(SP89b) while the results from the 1990-91 measurements are detailed in
Gaier et al., 1992 (SP91a) and Schuster et al., 1993 (SP91b). The results
from these measurements and the five balloon-borne Millimeter-wave
Anisotropy eXperiment (MAX) are summarized in Lubin, 1994. In order to
obtain additional data and frequency coverage, we returned to the
Amundsen-Scott South Pole Station during the austral summer 1993-94.

\section{Instrumentation}

The SP94 observations used the Advanced Cosmic Microwave Explorer (ACME) as
have all our previous degree scale anisotropy measurements. ACME is a one
meter off axis Gregorian telescope which has been described in detail in
SP89b. During the observations, the ellipsoidal secondary oscillated
sinusoidally at 8 Hz with a peak to peak throw of 3$^{\circ }$ on the sky.
The receiver signals were phase synchronously demodulated using a
``square-wave'' lockin amplifier and sampled every 0.5 seconds. The beam
profile of the telescope can be approximated as a Gaussian beam with a 1 $%
\sigma _b$ dispersion which varies in frequency as given below. Two
different total power radiometers were used in these observations. The lower
frequency (Ka-band) receiver is similar to that described in SP91a and
incorporates a very low noise, high electron mobility transistor (HEMT)
amplifier (Pospieszalski et al., 1990) cooled to 4 K in a $^4$He dewar. This
receiver operated at four center frequencies (27.25, 29.75, 32.25, and 34.75
GHz) with 2.5 GHz 3 dB bandwidths. The band subdivision is used to
compensate for gain variations across the full band and to obtain spectral
information which can be used to discriminate between the various
astrophysical foregrounds. For the Ka-band system, the beam dispersion is
given by $\sigma _b=0.^{\circ }70\pm 0.^{\circ }04\times \left( \frac{27.7}{%
\upsilon _{GHz}}\right) .$ The higher frequency receiver (Q-band) is
described in Gundersen et al., 1994 and also uses a cryogenic HEMT amplifier
based on a design developed at the National Radio Astronomy Observatory
(NRAO). This amplifier was built at UCSB with assistance from NRAO and uses
an AlInAs/GaInAs/InP HEMT (Pospieszalski et al., 1994) in the first of five
amplification stages. The Q-band system was multiplexed into 3 equal bands
centered at 39.15, 41.45, and 43.75 GHz with nominal 3 dB bandwidths of 2.3
GHz, and the beam dispersion is given by $\sigma _b=0.^{\circ }47\pm
0.^{\circ }04\times \left( \frac{41.5}{\upsilon _{GHz}}\right) .$ The HEMT
amplifiers introduce intrinsic cross correlations between the bands which
can be characterized by the correlation coefficient between any two
frequencies. The measured correlations were typically 0.25 and 0.50 for the
Ka and Q-band systems, respectively, including atmospheric correlations. The
radiometers are calibrated to 10\% absolute accuracy and 3\% relative
accuracy using a combination of cryogenic cold loads, the sky, ambient
Eccosorb, and the Moon. The long term stability of the system was checked
daily by inserting an ambient load ``calibrator''. These calibrations varied
by less than 3\% over the time scale of an observation and contribute a
negligible amount to the final error estimate.

\section{Observations}

Two observations were performed between January 9, 1994 and January 22, 1994
and collected 261 hours of data. The first observation used the Q-band
receiver and the second observation used the Ka-band receiver. These
observations consisted of smooth, constant declination, constant velocity
scans of length 20$^{\circ }$ on the sky about a center $\alpha _{cen}=$45$%
^{\circ }$, $\delta _{cen}=$-62$^{\circ }$. The closest approach to the Sun
was 60$^{\circ }$ on the sky and the closest approach to the plane of the
galaxy corresponds to $b^{II}=-40^{\circ }$, $l^{II}=272^{\circ }$. This is
a low foreground emission region which overlaps some of the region observed
in SP91. Our measurement of the Eta Carina region showed that our absolute
elevation was one degree lower than we expected. The offset has been
attributed to sag in the inner frame of the telescope mount and makes a
direct comparison between these measurements and the SP91 measurements
problematic. The instantaneous right ascension of the beam for any of the 3
observations can be given by $\alpha (t)=\alpha _{cen}+20\left( S\left(
t\right) mod\,2-1/2\right) /\cos \delta _{cen}+\alpha _o\sin \left( 2\pi \nu
_{ch}t\right) /\cos \delta _{cen}$ where $S\left( t\right) =int\left(
\upsilon _{sc}t+1\right) $ enumerates the scan number, $\alpha _o=1.5^{\circ
}$ is the sinusoidal chop amplitude, $\nu _{ch}=8$ Hz is the chop frequency,
and $\upsilon _{sc}=10$ mHz is the scan frequency. The instantaneous beam
position on the sky is then given by $\varphi \left( t\right) =\alpha
(t)\cos \delta _{cen}$. Observations of the Moon established the absolute
pointing at low elevations and this was confirmed with observations of Eta
Carina at high elevations. The error in absolute pointing is $\pm
0.25^{\circ }$ in right ascension and $\pm 0.12^{\circ }$ in declination
while the error in relative pointing is $\pm 0.05^{\circ }$ in right
ascension and $\pm 0.05^{\circ }$ in declination. If a temperature at
position $\widehat{n}_i^k$ is compared to a temperature at position $
\widehat{n}_j^l$ at the same declination, then the dimensionless window
function can be written as
\begin{equation}
\label{Eq1}W_\ell (\Phi _{ij}^{kl})=B_\ell (\sigma _b^k)B_\ell (\sigma _b^l)
\sum_{r=0}^{\ell}\frac{\left( 2\ell -2r\right) !\left( 2r\right) !}{\left[
2^\ell r!\left( \ell -r\right) !\right] ^2}4H_0^2[(\ell-2r)\alpha
_o]j_0^2[(\ell-2r)\Delta \varphi /2]\cos [(\ell -2r)\Phi _{ij}^{kl}]\
\end{equation}
where $\Phi _{ij}^{kl}=\cos {}^{-1}(\widehat{n}_i^k\cdot $ $\widehat{n}_j^l)$
is the angular difference (or lag) between the temperature measured at bin $%
i $ with channel $k$ and the temperature measured at bin $j$ with channel $l$%
. The beam profile function is given by $B_\ell (\sigma _b)=\exp \left[
-\ell \left( \ell +1\right) \sigma _b^2/2\right] ,$ $H_0$ is the Struve
function of 0 index, $j_0$ is the 0$^{th}$ order spherical Bessel function
and $\Delta \varphi =(20/43)(\pi /180)$ is the bin size in radians on the
sky. The indices are given by $i,j=1$ to $N=43$ bins and $k,l=1$ to $F=3 $
for the Q-band data and $k,l=1$ to $F=4$ for the Ka-band data. This window
function is shown in Figure 2 for the different combinations of beamsizes
and is a specific example taken from a more general expression in White and
Srednicki (1994).

\section{Data Reduction and Analysis}

Data were rejected for a number of reasons including poor pointing/chopper
performance (1.2\%), performance of other observations (0.2\%), calibration
sequences (0.7\%), telescope/receiver maintenance (9.1\%), temperature
variations of the cold plate and backend electronics (0.9\%), and bad
weather (12.9\%). From a total of 261 hours of data, 196 hours were used in
the data analysis. The poor weather data were determined in a way similar to
SP91b in which the data from a single scan were combined into position bins
from which an average and one sigma error bar were calculated. The $\chi ^2$
of the individual scans was then calculated, and if the probability of
exceeding $\chi ^2$ for 43 degrees of freedom was less than 0.01 for any
channel, then the data from {\it all} channels for that scan were removed.
As a cross check, other weather filters similar to SP91a were also
implemented with no significant changes in the final data set. As with SP89
and SP91, an offset and gradient were removed in time over the time scale of
a single scan (100 secs) for each of the channels in an observation. The
offsets were between 1 and 2 mK depending on the channel. The offset and
gradient subtraction are taken into account in the analysis by creating a
matrix $R$ such that $\widetilde{T}_a^k=R_{aj}^{kl}T_j^l$ where $T_j^l$ are
the $F\times N$ temperature means and $\widetilde{T}_a^k$ are the $F\times
(N-2)$ projected temperatures with $a=1,$ $N-2.$ The formation of $R$ is
discussed in Bunn et al. 1994 and we have made $R$ orthogonal such that $%
R^TR=I$. The coadded means and 1 sigma error bars are shown in Figure 1 and
show statistically significant, correlated signals for each observation. All
quoted temperatures have been converted from antenna temperature to
thermodynamic temperature and have been corrected for atmospheric
absorption. In order to determine the origin of the observed structure, the
data were binned in azimuthal and heliocentric coordinate systems. Various
subset analyses were performed including dividing the data set into four
roughly equal quarters in time and dividing the data into four roughly equal
quarters depending on the azimuthal position of the telescope beam. None of
these analyses suggest that the observed structure is anything but celestial
in origin.

\section{Astrophysical Foregrounds}

There are several astrophysical foregrounds which could contaminate these
observations. These include diffuse synchrotron and free-free emission from
within our own galaxy and extragalactic emission from discrete radio
sources. Neither the Sunyaev-Zeldovich effect nor diffuse 20 K dust emission
is expected to contribute more than a few $\mu $K signal. If we assume that
the 408 MHz map (Haslam et al.) is a tracer of diffuse, high galactic
latitude synchrotron emission, then we calculate the rms differential
synchrotron emission to be 1.0 K at 408 MHz for these observations. Given
the spectral index for diffuse synchrotron is $\beta =-2.8$ with an antenna
temperature given by $T_A\propto \upsilon ^\beta $, we estimate that the
diffuse synchrotron contribution to the observed rms to be $<7$ $\mu $K in
the lowest frequency channel and $<3$ $\mu $K in the highest frequency
channel. The small amount of differential emission that exists in this
region of the sky at 408 MHz can be correlated with discrete radio sources
which have been observed in other source surveys at 408 MHz and are
identified in the PKSCAT90 database (Wright and Otrupcek, 1990). Unlike
diffuse synchrotron emission, discrete radio sources and free-free emission
cannot be dismissed on amplitude arguments alone. For diffuse free-free
emission, there are no all-sky surveys which would allow a direct estimate
of the free-free contribution to the observed structure. Instead, we have to
rely on estimates based on a 10$^{\circ }$ $\times $ 12$^{\circ }$ H$\alpha $
map (Reynolds, 1992) made at a similar angular scale to predict an average
background free-free brightness temperature that can be expressed as T$%
_{ff}=1\times 10^{-2}\upsilon _{GHz}^{-2.1}\csc \left| b^{II}\right| $.
Reynolds' data suggests that the variations in the H$\alpha $ intensity may
be a factor of 2 above the average, such that $\Delta $T$_{ff}=2$T$_{ff}$ .
{}From this we calculate the differential brightness temperature due to
free-free emission (upon the closest approach to the galactic plane) to
range from $\Delta $T$_{ff}=30$ $\mu $K at the lowest frequency to $\Delta $T%
$_{ff}=12$ $\mu $K at the highest frequency. There have been many discrete
source surveys at lower frequencies which are compiled in the PKSCAT90
database. The Parkes-MIT-NRAO (PMN) survey at 4.85 GHz (Wright et al., 1994)
serves as the most sensitive survey with a flux limit of 30 mJy in our
observation region and includes all the sources identified in PKSCAT90 for
our region. The sensitivity of the telescope is 90 $\mu $K/Jy (assuming
100\% aperture efficiency), so a 30 mJy flat spectrum source (with a flux
density S(Jy)$\propto \nu ^{\beta +2}\propto T_A\nu ^2,$ $\beta =-2$$)$
would produce a 3 $\mu $K signal, which is well below the noise of the
observations. Since the effective solid angle, $\Omega _e$, of the telescope
varies like $\Omega _e\propto \nu ^{-2}$, a flat spectrum point source with
a solid angle $\Omega _p\ll \Omega _e$ and $T_{flat}\propto \upsilon ^{-2}$
would give $T_A=\Omega _p\frac{T_{flat}}{\Omega _e}\propto \nu ^0$, while a
thermal point source produces an antenna temperature $T_A\propto \nu ^2$. If
we make the worst case assumption that all the point sources have flat spectra
to
45 GHz, then we estimate that they would produce a $\Delta $T$_{rms}=30-45$ $%
\mu $K. Since the worst case estimates of contamination from free-free
emission and flat spectrum point sources are comparable to the rms level of the
observed structure, we cannot dismiss or verify these types of foreground
contamination without measuring $\beta .$ This is addressed in the following
likelihood analysis.

\section{Likelihood Analysis}

We use the Bayesian method with a uniform prior (Bond et al. 1991) in the
determination of the root mean square (rms) amplitude of the data and to
make an estimate of the broad-band power in the CMB power spectrum (Bond,
1994a, and Steinhardt, 1994). The experimental two point correlation
function is given by %\begin{equation}
\begin{equation}
\label{Eq2}C(\widehat{n}_i^k\cdot \ \widehat{n}_j^l)=\left\langle \frac{%
\Delta T(\widehat{n}_i^k)\Delta T(\widehat{n}_j^l)}{T_0^2}\right\rangle
=\frac 1{4\pi }\sum_{\ell =2}^\infty (2\ell +1)C_\ell W_\ell (\Phi
_{ij}^{kl})
\end{equation}
where $\left\langle a_{\ell m}a_{\ell ^{^{\prime }}m^{^{\prime
}}}\right\rangle =\delta _{\ell \ell ^{^{\prime }}}\delta _{mm^{^{\prime
}}}C_\ell $ for a spherical harmonic expansion of the radiation temperature
given by $\frac{\Delta T(\widehat{n})}{T_0}=\sum_{\ell m}a_{\ell m}Y_{\ell
m}(\widehat{n})$, $W_\ell $ is the window function (Eq. 1) and $T_0=2.726\pm
0.01$ $K$ (Mather et al. 1994). The rms amplitude is calculated from
\begin{equation}
\label{Eq3}\left( \frac{\Delta T}{T_0}\right) _{rms}^2=\frac 1{4\pi
}\sum_{\ell =2}^\infty (2\ell +1)C_\ell \overline{W}_\ell ,\qquad \overline{W%
}_\ell =\frac 1{N_{pix}}\sum_{ikl}W_\ell (\Phi _{ii}^{kl})
\end{equation}
where $N_{pix}=F^2\times N$ and $\overline{W}_\ell $ is the average window
function at zero lag. Following Bond, 1994a, the broad band power estimate
is given by
\begin{equation}
\label{Eq4}\left\langle \overline{C}_\ell \right\rangle _B=\left\langle
\frac{C_\ell \ell (\ell +1)}{2\pi }\right\rangle _B=\frac{\left( \Delta
T/T_0\right) _{rms}^2}{\sum_{\ell =2}^\infty (\ell +\frac 12)\overline{W}%
_\ell /(\ell (\ell +1))}.
\end{equation}
We consider a scale invariant, $n=1,$ ``flat'' radiation power spectrum
given by $C_\ell \propto (f^kf^l)^\beta /(\ell (\ell +1))$ where the
constant of proportionality and $\beta $ are determined in the likelihood
analysis and $f^k$ is the center frequency of channel k normalized to the
lowest center frequency of the observation(s). The full covariance matrix is
given by $M_{ij}^{kl}=C(\widehat{n}_i^k\cdot \ \widehat{n}%
_j^l)+D_{ij}^{kl}\sigma _i^k\sigma _j^l/T_0^2$ where $D_{ij}^{kl}\sigma
_i^k\sigma _j^l$ is the data covariance matrix which is diagonal in each of
the $F^2$ $N\times N$ submatrices. This covariance matrix accounts for all
spatial and channel-channel correlations as well as the error for each bin
at each frequency. The offset and gradient subtraction are taken into
account by creating a projected covariance matrix given by $\widetilde{M}%
=RMR^T$, where $R$ is defined in Section 4. The likelihood of the data set
is proportional to $\left| \widetilde{M}\right| ^{-1/2}\exp (-\chi ^2/2)$,
where $\chi ^2= \widetilde{T}^T\widetilde{M}^{-1}\widetilde{T}$ for $F\times
(N-2)$ degrees of freedom. Table 1 lists the resulting rms and band power
estimates for the individual observations as well as the combined
observations. Figure 2 shows the band power estimates with $\beta =0$ in
relation to the COBE band power estimate.

\section{Discussion}

The Q-band broad band power is consistent with standard CDM normalized to
COBE and has a spectral index, $\beta $, which is consistent with the CMB
spectrum. The Ka-band observation is also consistent with standard CDM
normalized to COBE; however, the spectral index does not rule out discrete
radio sources and thus does not afford such a straightforward
interpretation. When the Ka and Q-band data are combined, the broad band
power is consistent with the standard CDM model normalized to COBE and the
spectral index is consistent with the CMB spectrum. For the standard CDM
power spectrum, the amplitude of the combined Ka+Q observations corresponds
to a $Q_{rms-PS}=16.2_{-2.6}^{+5.4}$ $\mu $K. The combined Ka+Q data show a
marked improvement on the 1 $\sigma $ confidence interval for the best fit $%
\beta .$ For the most probable $\Delta T_{rms}$, spectra with $\beta <-2$
(such as diffuse synchrotron and free-free emission) are formally excluded
at the 5 $\sigma $ level. The SP94 results are most easily compared to the
SP91 results and the results of Wollack et al. 1993 (SK93). A reanalysis
(Bond, 1994b) of the combined SP91a and SP91b observations gives $%
\left\langle \overline{C}_\ell \right\rangle _B=1.06_{-0.27}^{+0.83}\times
10^{-10}$ which compares to the Ka-band SP94 results of $\left\langle
\overline{C}_\ell \right\rangle _B=1.17_{-0.42}^{+1.33}\times 10^{-10}$. The
Bond, SP91 analysis assumes that $\beta =0$ and that there is no channel to
channel correlations. The SK93 result gives $\left\langle \overline{C}_\ell
\right\rangle _B=1.31_{-0.7}^{+1.2}\times 10^{-10}$, for $\beta
=-0.3_{-1.2}^{+0.7}$. The SK93 result is consistent with both the SP94 and
the combined SP91 results, although one should note the differences between
the experiments which are addressed in SK93.

\acknowledgments

We would like to thank M. Pospieszalski, M. Balister, and W. Lakatosh of
NRAO-CDL for useful information regarding amplifiers and for supplying the
26-36 GHz HEMT amplifier. In addition we would like to thank L. Nguyen of
Hughes Research Labs for providing the InP HEMTs which we've incorporated
into our 38-45 GHz amplifiers. This project would have been impossible
without the support of B. Sadoulet and the Center for Particle Astrophysics.
D. Fischer and the whole Antarctic Support Associates staff provided the
valuable support and expertise at the Amundsen-Scott South Pole station. We
are grateful to N. Sugiyama for providing us with the CDM\ radiation power
spectrum and to R. Bond, M. Srednicki, P. Steinhardt, M. White, and L. Page
for useful discussions regarding data analysis and window functions. We
would like to acknowledge the previous contributions of J. Schuster. N.
Figueiredo is partially supported by Conselho Nacional de Desenvolvimento
Cientifico e Tecnologico, Brazil. This work was supported by NSF grant OPP
92-21468 and AST 91-20005.

The means, covariances and window functions will be made available from
rot.ucsb.edu using anonymous FTP in the directory [sp94dat] and from the Web
site http://www.deepspace.ucsb.edu.

\clearpage

\makeatletter
\def\jnl@aj{AJ} \ifx\revtex@jnl\jnl@aj\let\tablebreak=\nl\fi
\makeatother

\begin{planotable}{clrcc}
\tablewidth{23pc}
\tablecaption{Derived Parameters From Likelihood Analysis $(\pm 1\sigma)$}%From
%%Likelihood Analysis Using $C^{flat}_\ell$}
\tablehead{
\colhead{Band} & \colhead{$\Delta T_{rms}$} &
\colhead{$\beta$} & \colhead{$\Delta T_{rms}$} &
\colhead{$\left\langle \frac{C_\ell\ell(\ell+1)}{2\pi} \right\rangle_B$}
\\[.2ex]
\colhead{} & \colhead{($\mu$K)} &
\colhead{} & \colhead{($\mu$K)} &
\colhead{$\times 10^{-10}$} \\[.2ex]
\colhead{(1)} & \colhead{(2)} &
\colhead{(3)} & \colhead{(4)} &
\colhead{(5)}}

\startdata
Q-band & $42.8^{+16.9}_{-7.3}$ & $1.7^{+1.5}_{-1.6}$ & $49.3^{+19.3}_{-8.3}$
& $2.14^{+2.00}_{-0.66}$ \nl
Ka-band & $30.8^{+14.7}_{-6.1}$ & $0.2^{+0.9}_{-1.4}$ & $32.0^{+14.7}_{-6.5}$
& $1.17^{+1.33}_{-0.42}$ \nl
Ka+Q & $31.9^{+11.6}_{-6.1}$ & $0.9^{+0.3}_{-0.6}$ & $42.0^{+15.8}_{-6.8}$
& $1.77^{+1.58}_{-0.53}$ \nl

\tablecomments{(cols. [2-3]) The rms amplitude and spectral index $\beta$
determined from the likelihood analysis using a scale invariant, n=1, ``flat"
radiation power spectrum given by $C_\ell\propto(f^kf^l)^\beta/(\ell(\ell+1))$,
where
$f^k$ is the center frequency of channel k normalized to the lowest
frequency of the observation; (col. [4]) the
rms amplitude for $\beta=0$; (col. [5]) CMB broad-band power spectrum estimate
for $\beta=0$.}
\end{planotable}

\clearpage

\begin{figure}
\centerline{\epsfxsize=5in\epsfbox{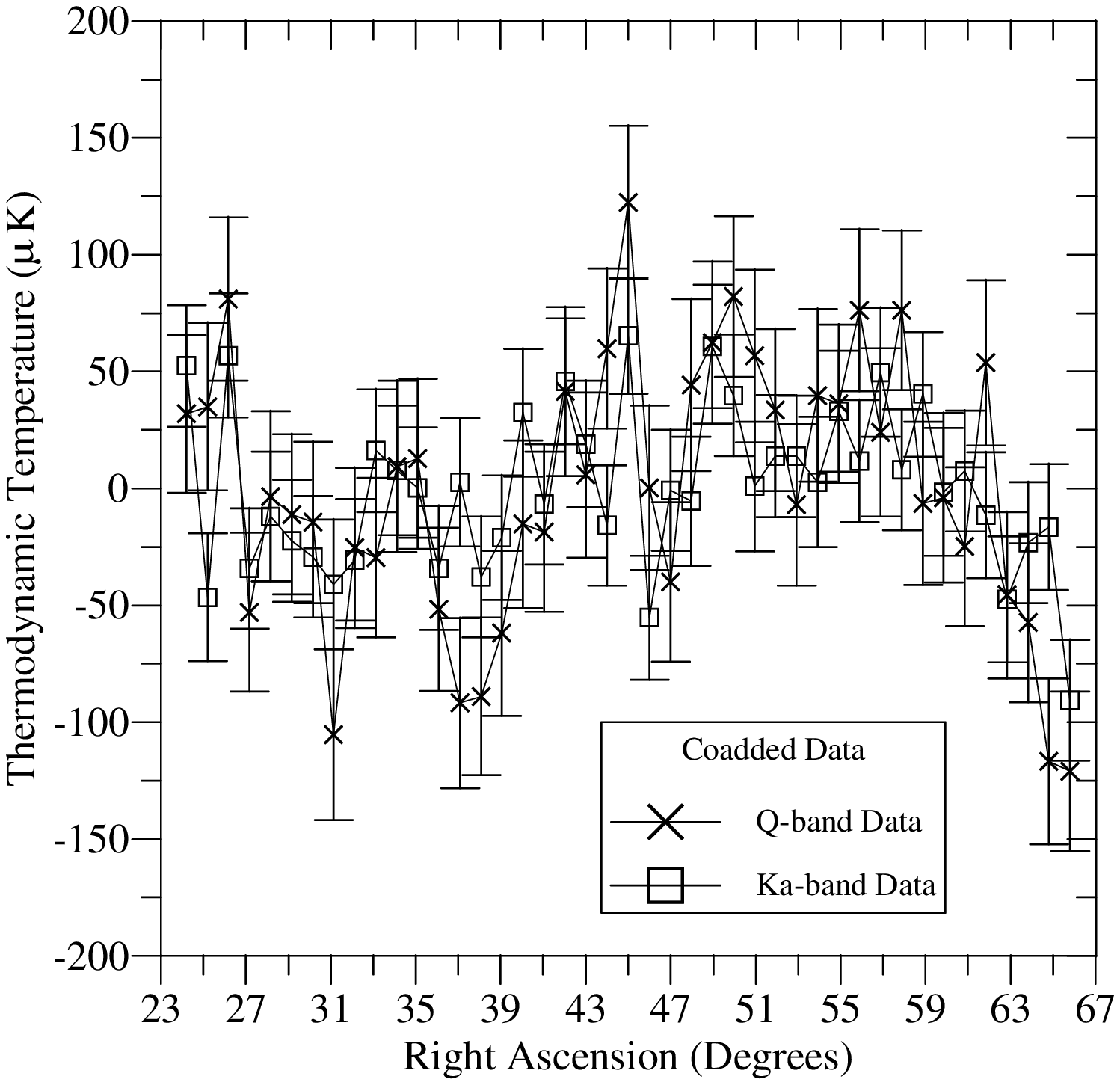}}
\caption{Coadded single difference data in $\mu $K thermodynamic temperature,
referenced to the top of the atmosphere with $\pm 1$ $\sigma $ error bars. The
error bars account for the correlated noise between the channels. The data in
this figure
shows that there is significant correlated structure. This data set was not
explicitly used
in the analysis since the spatial correlations between the bins have to be
taken into account
using the full theoretical covariance matrix.}
\end{figure}

\begin{figure}
\centerline{\epsfxsize=5in\epsfbox{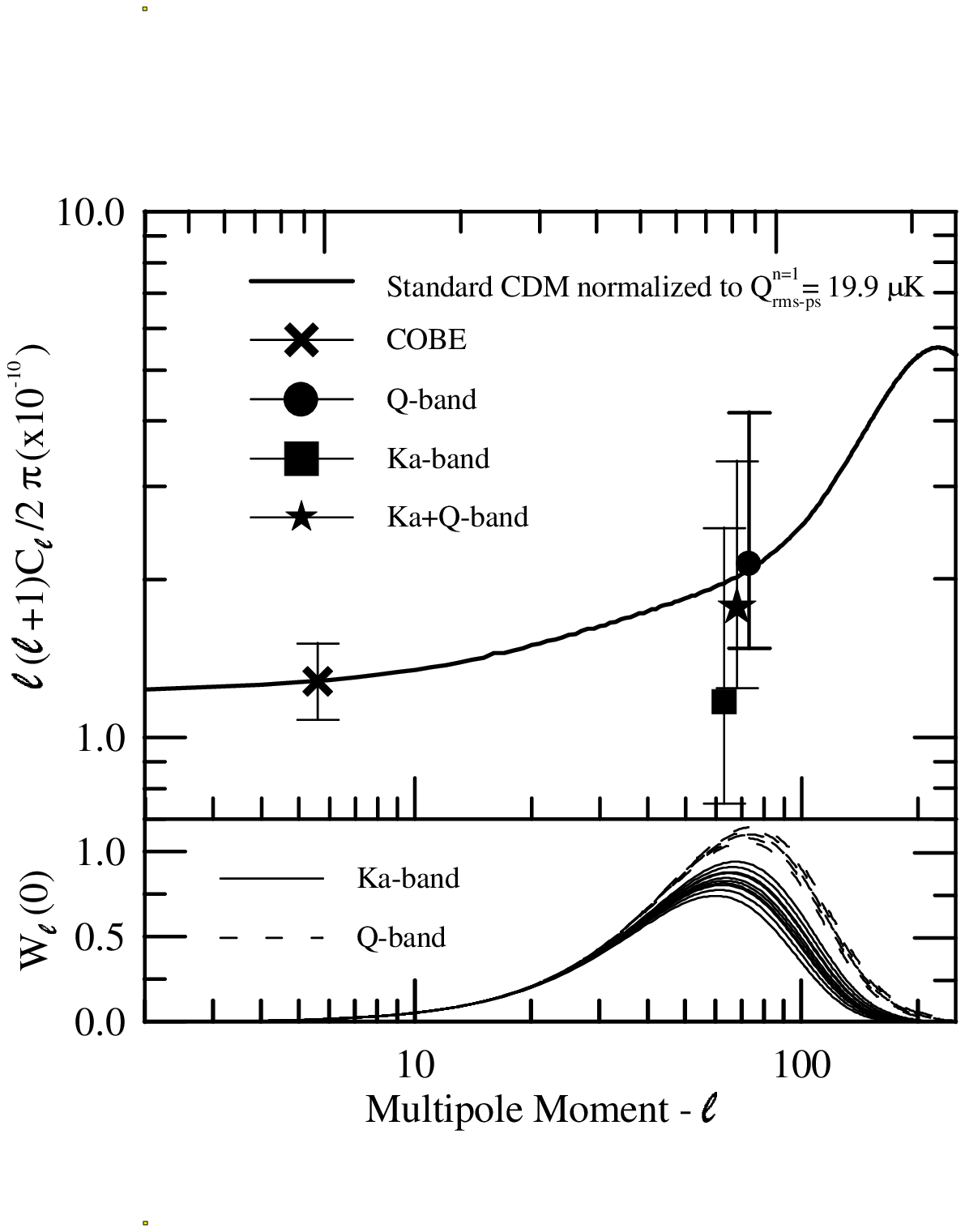}}
\caption{The top figure shows the band power estimates as given in Table
1 compared to the standard CDM radiation power spectrum (Sugiyama, 1994)
normalized to the COBE
$Q_{rms-ps}^{n=1}=19.9 \pm 1.6$ $\mu K$ (Gorski et al., 1994).  The lower
figure shows the window function at zero lag for the Q and Ka-band receivers.
The
different traces represent the window functions for the various combinations
of beamsizes for each receiver. The combined Ka+Q-band window function
is not shown for clarity sake.}
\end{figure}

\end{document}